%% file: IFE_TM.tex
\documentclass[
    aps,
    prb,
    twocolumn,
    floatfix,nofootinbib,
    superscriptaddress
]{revtex4-2}

\usepackage{graphicx}
\usepackage{subfigure}
\usepackage[utf8]{inputenc}
\usepackage{times}
\usepackage[breaklinks,colorlinks=true]{hyperref}
\hypersetup{linkcolor=red,citecolor=green,urlcolor=blue}
\usepackage{color}
\usepackage{siunitx}
\usepackage{booktabs}
\usepackage[version=4]{mhchem}
\usepackage{ulem}
\usepackage{threeparttable}

\usepackage{amssymb} 
\usepackage{amsmath}
\usepackage{amsthm}
\usepackage{braket}
\usepackage{MnSymbol}
\usepackage{stmaryrd}
\usepackage{mathtools}

\input{commands}
\usepackage{tikz}
\usepackage{tikz-cd}
\usetikzlibrary{calc}
\usetikzlibrary{decorations.markings}
\usepackage{xcolor}
\usepackage{graphicx}

\makeatletter
\renewcommand\@biblabel[1]{#1.}
\makeatother
\begin{document}

\setcounter{secnumdepth}{2} 

\title{Light-induced Orbital and Spin Magnetism in $3d$, $4d$, and $5d$ Transition Metals}

\author{Theodoros Adamantopoulos}
    \thanks{t.adamantopoulos@fz-juelich.de}
    \affiliation{\mainz}
    \affiliation{\pgi}
    \affiliation{\aachen}



\author{Dongwook Go}
    \affiliation{\mainz}
    
\author{Peter M. Oppeneer}
    \affiliation{\uppsalaa}
    \affiliation{\uppsalab}
    

\author{Yuriy Mokrousov}
    \affiliation{\pgi}
    \affiliation{\mainz}

\date{\today}

\begin{abstract}
Understanding the coherent interplay of light with the magnetization in metals has been a long-standing problem in ultrafast magnetism. While it is known that when laser light acts on a metal it can induce magnetization via the process known as the inverse Faraday effect (IFE), the most basic ingredients of this phenomenon are still largely unexplored. In particular, given a strong recent interest in orbital non-equilibrium dynamics and its role in mediating THz emission in transition metals, the exploration of distinct features in spin and orbital IFE is pertinent. Here, we present a first complete study of the spin and orbital IFE in $3d$, $4d$ and $5d$ transition metals of groups IV$-$XI from first-principles. By examining the dependence on the light polarization and frequency, we show that the laser-induced spin and orbital moments may vary significantly both in magnitude and sign.
We underpin the interplay between the crystal field splitting and spin-orbit interaction as the key factor which determines the magnitude and key differences between the spin and orbital response.
Additionally, we highlight the anisotropy of the effect with respect to the ferromagnetic magnetization and to the crystal structure. The provided complete map of IFE in transition metals is a key reference point in the field of optical magnetism.

\end{abstract}

\maketitle






\date{\today}


%
%
%
%
%
%
%
%
%
%
%
%
%
%
%

\noindent {\bf Introduction} 

\noindent The demonstration of ultrafast demagnetization in ferromagnets by the application of femtosecond laser pulses~\cite{Beaurepaire_1996} gave rise to the field of ultrafast spintronics and set the stage for efficient manipulation of magnetism by light. By now it has been rigorously   demonstrated that all-optical helicity-dependent magnetization switching can be achieved in wide classes of magnetic materials~\cite{Kimel_2005,Stanciu_2007, Vahaplar_2009, Kirilyuk_2010, Steil_2011, Khorsand_2012, Alebrand_2012, Mangin_2014, John_2017}, thus paving the way to contactless ultrafast magnetic recording and information processing. In interpretation of the switching experiments, the inverse Faraday effect (IFE) $-$ i.e. the phenomenon of magnetization induced by a coherent interaction with light acting on a material $-$ has been considered as one of the major underlying mechanisms for the magnetization reversal. Although  theoretically predicted and experimentally observed decades ago~\cite{Pitaevskii_1960, Ziel_1965, Pershan_1966}, a consensus in the theoretical understanding of IFE is still lacking~\cite{Hertel_2006, Zhang_2009, Popova_2011, Popova_2012, Battiato_2014, Hertel_2015} -- however, several microscopic methods have been recently developed for the calculation of this phenomenon  in diverse setups~\cite{Berritta_2016, freimuth_2016, Qaiumzadeh_2016, Mironov_2021, Banerjee_2022, Zhou_2022, Sharma_2024, Wong_2024}.

Apart from its established role in the magnetization switching of ferromagnets and ferrimagnents, the impact of IFE in the THz regime has also been studied in antiferromagnets like CrPt~\cite{Dannegger_2021} and Mn$_2$Au~\cite{Merte_Mn2Au}. Moreover, it has been shown that the component of IFE which is perpendicular to the magnetization direction behaves differently with respect to the light helicity than the parallel component~\cite{freimuth_2016}, giving rise to helicity-dependent optical torques~\cite{Choi_2017, Iihama_2021, Iihama_2022} and THz emission~\cite{Huisman_2016}. The impact of the optical torques has been demonstrated to be significant in antiferromagnetic Mn$_2$Au~\cite{Freimuth_2021_Mn2Au} where transverse IFE may be responsible for the THz emission~\cite{Behovits_2023, Huang_2024}, provide an alternative way to switching the magnetization~\cite{Ross_2024} as well as to drive domain wall motion~\cite{Gavriloaea_2024}. Recently it has also been shown that light can induce colossal magnetic moments in altermagnets~\cite{Adamantopoulos_npj_2024}. Although IFE is conventionally associated with an excitation by circularly polarized light, it can also be activated by linearly-polarized laser pulses~\cite{Adamantopoulos_npj_2024, Merte_Mn2Au, Zhou_2022, Ali_2010, Gridnev_2013, Mrudul_2024}. 

In solids, two contributions to the magnetization exist: due to spin and orbital moment of electrons. And while the discussion of IFE is normally restricted to the response of spin, since recently, non-equilibrium dynamics of orbital angular momentum started to attract significant attention~\cite{Go_2021b, Go_2024_Europhysics, Jo_2024}. Namely, the emergence of orbital currents in the context of the orbital Hall effect~\cite{Tanaka_2008, Kontani_2009, Go_2018, Choi_2023}, orbital nature of current-induced torques on the magnetization~\cite{Go_2020a, Go_2020b, Lee_2021}, and current-induced orbital accumulation sizeable even in light materials~\cite{dgo_srep, Salemi_2019, Go_2021a, Lyalin_2023} have been addressed theoretically and demonstrated experimentally. Notably, in the context of light-induced magnetism, Berritta and co-workers have predicted that the IFE in selected transition metals can exhibit a sizeable orbital component~\cite{Berritta_2016}, with the generality of this observation reaching even into the realm of altermagnets~\cite{Adamantopoulos_npj_2024}. At the same time it is also known that within the context of plasmonic IFE the orbital magnetic moment due to electrons excited by laser pulses in small nanoparticles of noble and simple metals can reach atomic values~\cite{Hurst_2018, Cheng_2020, Lian_2024}. Despite the fact that very little is known about the interplay of spin and orbital IFE in real materials, it is believed that IFE is a very promising  effect in the context of orbitronics $-$ a field which deals with manipulation of the orbital degree of freedom by external perturbations.

\begin{figure*}[t!]
\begin{center}
\rotatebox{0}{\includegraphics [width=0.85\linewidth]{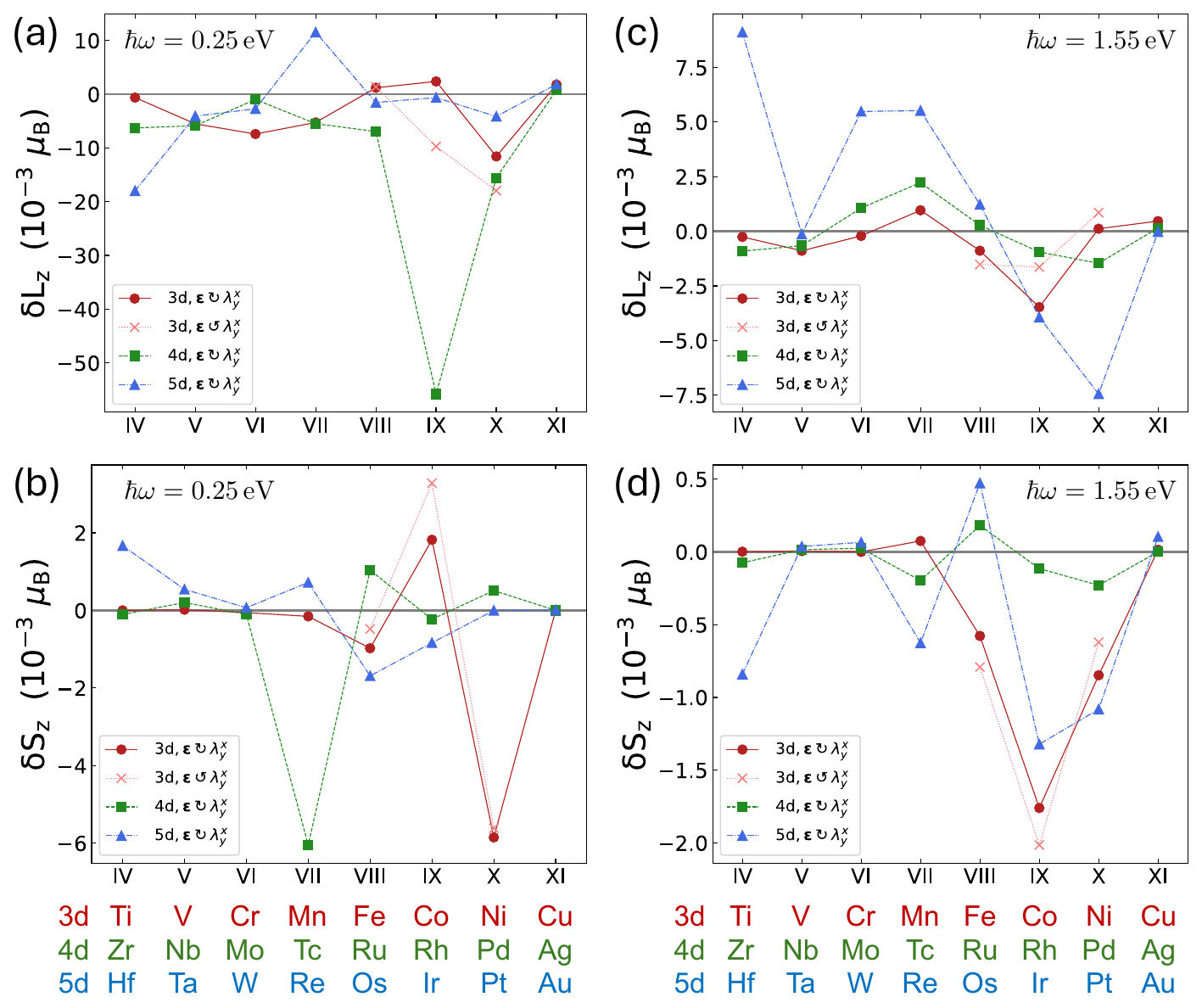}}
\end{center}
\caption{{\bf Map of light-induced magnetism in transition metals}. (a-d) Light-induced orbital $\delta L_z$ (a, c) and spin $\delta S_z$ (b, d) magnetic moments in $3d$ (red circles), $4d$ (green squares) and $5d$ (blue triangles) transition metals of groups IV$-$XI. In all calculations light is considered to be circularly polarized in the $xy$-plane. For the $3d$ magnetic elements (Fe, Co, Ni) the moments that arise for left-handedly polarized light are shown with light red crosses. The light frequency is $\hbar\omega=0.25$\,eV in (a-b) and $\hbar\omega=1.55$\,eV in (c-d).}
\label{Fig1}
\end{figure*}

Although the  importance of IFE in mediating the light-induced magnetism  is beyond doubt, a comprehensive in-depth material-specific knowledge acquired from microscopic theoretical analysis of this effect is still missing. While acquiring this knowledge is imperative for the field of ultrafast spintronics, since the initial seminal work by Berritta and co-workers~\cite{Berritta_2016}, very little effort has been dedicated to the  explorations of this phenomenon from first principles. Here, we fill this gap by providing a detailed first-principles study of the light-induced spin and orbital magnetism for the $3d$, $4d$ and $5d$ transition metals of groups IV$-$XI. By exploring the dependence on the frequency and the polarization of light, we acquire insights into the origin and differences between spin and orbital flavors of IFE, as well as key features of their behavior. Our work provide a solid foundation for further advances in the field of optical magnetism and  interaction of light and matter.
\vspace{0.2cm}

\noindent {\bf Results} 

\noindent
{\bf Light-induced magnetism in transition metals}


\noindent
We begin our discussion by exploring the IFE in a series of $3d$, $4d$ and $5d$ transition metals of groups IV$-$XI. IFE is a non-linear optomagnetic effect in which magnetization $\delta\mathcal{}{O}\propto\mathbf{E}\times \mathbf{E}^*$ is induced as a second-order response to the electric field $\mathbf{E}$ of a laser pulse. We calculate the light-induced spin $\delta S$ and orbital $\delta L$ magnetic moments by means of the Keldysh formalism, see the section Methods~\cite{freimuth_2016,Adamantopoulos_npj_2024}. We focus on light energies $\hbar\omega$ of 0.25\,eV and 1.55\,eV which have been routinely used before in our studies within the same formalism~\cite{Merte_Mn2Au, Adamantopoulos_2024, Adamantopoulos_npj_2024}. Additionally, we choose a lifetime broadening $\Gamma$ of 25\,meV that corresponds to room temperature in order to account for effects of disorder on the electronic states and to provide realistic estimations of the effect. In all calculations we choose the light to be circularly polarized in the $xy$ or $yz$ planes with a light intensity $I$ of 10\,GW/cm$^2$, while for magnetic materials  ferromagnetic magnetization is considered to be along the $z$-axis.

\begin{figure*}[t!]
\begin{center}
\rotatebox{0}{\includegraphics [width=0.85\linewidth]{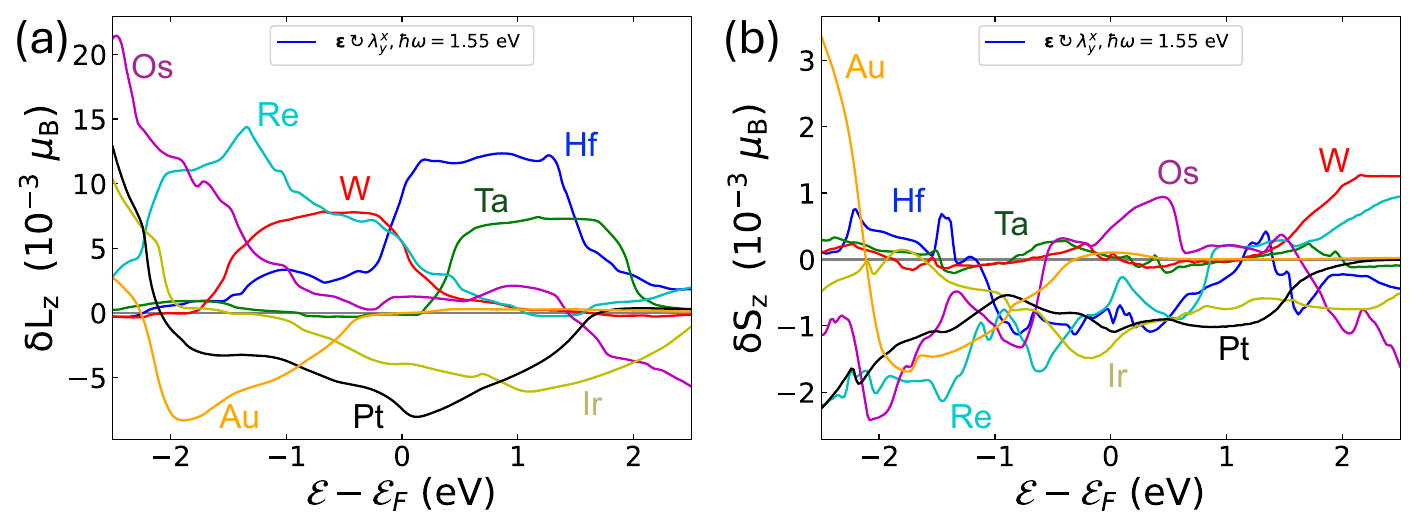}}
\end{center}
\caption{{\bf Band filling dependence of light-induced magnetic moments in $\mathbf{5d}$ transition metals}. (a-b) Light-induced orbital $\delta L_z$ (a) and spin $\delta S_z$ (b) magnetic moments in relation to the band filling for the $5d$ transition metals of groups IV$-$XI. Light is circularly polarized in the $xy$-plane and the light frequency is set at $\hbar\omega=1.55$\,eV.}
\label{Fig2}
\end{figure*}

In Fig.~\ref{Fig1} we present the calculated light-induced $\delta L_z$ [Fig.~\ref{Fig1}(a-c)] and $\delta S_z$ [Fig.~\ref{Fig1}(b-d)] moments of the considered transition metals at the Fermi energy. Light is circularly polarized in the $xy$ plane, with energies $\hbar\omega$ of 0.25\,eV [Fig.~\ref{Fig1}(a-b)] and 1.55\,eV [Fig.~\ref{Fig1}(c-d)]. Regarding the $3d$ magnetic elements Fe, Co, Ni, we analyze both right and left handed circular polarizations of light. The exact values of $\delta L_z$ and $\delta S_z$ are listed in Tables~\ref{table_1} and~\ref{table_2} of the Supplemental Material (SM). 
Not shown are computed transverse $x$ and $y$ components of the light-induced moments, which we find to be orders of magnitude smaller for the case of $xy$-polarized light, and which we discuss later.

At first sight we notice a strong dependence of the magnitude and sign of both $\delta L_z$ and $\delta S_z$ on the light frequency. We point out the case of fcc Rh where an colossal $\delta L_z=-56\cdot10^{-3}\mu_{\mathrm{B}}$ is predicted at $\hbar\omega=0.25$\,eV and which is drastically reduced at the higher frequency, while the spin component remains suppressed in both cases. For the case of magnetic hcp Co we notice a helicity-dependent change of sign for $\delta L_z$ at $\hbar\omega=0.25$\,eV which is not present for $\delta S_z$. In general, $\delta L_z$ varies stronger with the light helicity than $\delta S_z$. Moreover, $\delta L_z$ is one to two orders of magnitude larger than $\delta S_z$ for the non-magnetic elements, while they are of the same magnitude for the magnetic elemental materials. The last two observations are in agreement with the findings of Ref.~\cite{Berritta_2016} and indicate how the spontaneous magnetization strongly influences the effect in ferromagnets by the time-reversal symmetry breaking. For comparison, in Table~\ref{table_Oppeneer} we list the values of the total light-induced moments, defined as the sum of $\delta L_z$ and $\delta S_z$, for the transitional metals studied in~\cite{Berritta_2016} with $\hbar\omega=1.55$\,eV. Our calculated values are of the same order of magnitude, with the exceptions of Au, and of Co for the case of left-handed polarization, which can be attributed to the difference in the computational methods.

\begin{table}[t!]
    \centering
    \begin{tabular}{c c}
    \hline
    \multicolumn{2}{c}{$\hbar\omega=1.55$\,eV} \\
    \hline
   & \\
     Material & $\delta L_z + \delta S_z (10^{-3}\,\mu_{\mathrm{B}}/\text{unit cell})$ \\ [0.5ex]
     \hline \\
     bcc Fe & $-$1.5 / $-$2.3 \\ [0.5ex]
     hcp Co & $-$5.3 / $-$3.6 \\ [0.5ex]
     fcc Ni & $-$0.7 /  0.2 \\ [0.5ex]
     fcc Cu &  0.51 \\ [0.5ex]
     fcc Pd & $-$1.7  \\ [0.5ex]
     fcc Pt & $-$8.5  \\ [0.5ex]
     fcc Au &  0.07 \\ [0.5ex]
    \hline\hline
    \end{tabular}
    \caption{Total light-induced magnetic moments for transition metals previously studied in~\cite{Berritta_2016} in units of $10^{-3}\,\mu_{\mathrm{B}}$ per unit cell. Light is circularly polarized in the $xy$-plane. For the $3d$ magnetic elements (Fe, Co, Ni) the moments that arise for both right/left-handedly polarized light are listed. The light frequency is $\hbar\omega=1.55$\,eV.}
    \label{table_Oppeneer}
\end{table}

From Fig.~\ref{Fig1}(c-d) we get a clear picture of how the induced moments scale with the strength of spin-orbit coupling (SOC). The effect is overall larger for the heavier $5d$ elements, both in the orbital $\delta L_z$ and spin $\delta S_z$ channels. However, at the lower frequency of $\hbar\omega=0.25$\,eV it is more difficult to draw such a conclusion since the corresponding photon energy falls into the range of SOC strength, which promotes the role of electronic transitions among spin-orbit split bands occurring within a narrow range around the Fermi energy and limited regions in $k$-space. In contrast, the use of a much larger frequency involves transitions among manifestly orbitally-distinct states which take place over larger portions of the reciprocal space with a more uniform impact of the spin-orbit strength on the magnitude of the transition probabilities.
From this discussion we have to exclude the ferromagnetic elements since the spontaneous magnetization induces bands splittings on the scale of exchange strength therefore making the effect much more complex.

\begin{figure*}[t!]
\begin{center}
\rotatebox{0}{\includegraphics [width=0.85\linewidth]{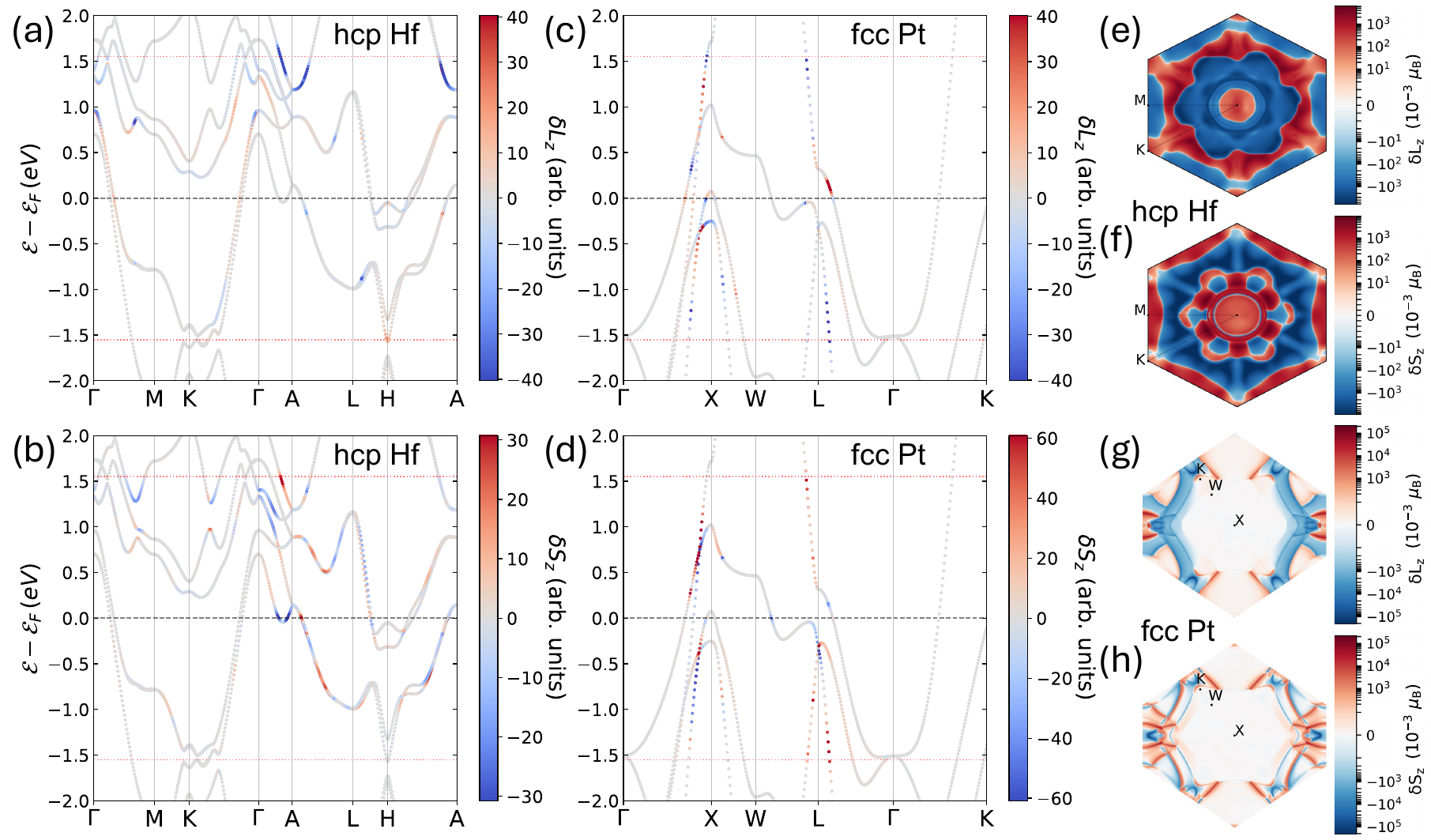}}
\end{center}
\caption{{\bf Reciprocal space anatomy of light-induced magnetism in nonmagnetic transition metals}. (a-d) Band-resolved light-induced orbital $\delta L_z$ (a, c) and spin $\delta S_z$ (b, d) magnetic moments in hcp Hf (a-b) and fcc Pt (c-d) relatively to the Fermi energy level $\mathcal{E}_{F}$ (dashed gray line). The horizontal dotted red lines at $\mathcal{E}_{F}\pm1.55$\,eV correspond to the frequency of the light. (e-h) Reciprocal space distribution of light-induced orbital (e, g) and spin (f, h) magnetic moments in hcp Hf (e-f) and fcc Pt (g-h) at the Fermi energy. In all calculations light is circularly polarized in the $xy$-plane at the frequency of $\hbar\omega=1.55$\,eV.}
\label{Fig5}
\end{figure*}

We study the dependence on the light polarization by performing additional calculations for light circularly polarized in the $yz$-plane, presenting the results for the $z$ and $x$ components of $\delta L$ and $\delta S$ respectively in Figs.~\ref{Fig1_suppl} and~\ref{Fig2_suppl} of the SM for $\hbar\omega=0.25$\,eV and $\hbar\omega=1.55$\,eV. Our results for $\delta S_z$ and $\delta S_x$ for the magnetic transition metals reproduce exactly the values of Ref.~\cite{freimuth_2016}, computed with the same method. We further present the values of $\delta L_z$ and $\delta L_x$, where we find the orbital response along the magnetization, $\delta L_z$, to be by a factor of two larger than the  corresponding spin response, when averaged over all magnetic elements. Generally, for the $z$-component the effect at the magnetic elements is similar in magnitude with the case of polarization in the $xy$-plane shown in Fig.~\ref{Fig1}, and even in the light helicity, while it vanishes for the non-magnetic elements. On the other hand, for the $x$-component the effect at the magnetic elements becomes one-two orders of magnitude smaller and is odd in the light helicity, although for the non-magnetic elements the values are  comparable to the case shown in Fig.~\ref{Fig1}. It is important to note that for light circularly-polarized in the plane containing  ferromagnetic magnetization, the light-induced transverse to magnetization induced moments, even though being smaller than the longitudinal ones, are related to light-induced torques that are experimentally demonstrated to lead to helicity-dependent THz emission~\cite{Huisman_2016, freimuth_2016}.

In order to get a better understanding of the impact of the time-reversal symmetry breaking on IFE, we consider the cases of ferromagnetic, non-relativistic (i.e. computed without SOC) ferromagnetic, and antiferromagnetic fcc Ni. In Figs.~\ref{Fig5_suppl} and~\ref{Fig6_suppl} of the SM we present the band filling dependence of the Cartesian components of the light-induced orbital moments $\delta L$, for $\hbar\omega=0.25$\,eV and $\hbar\omega=1.55$\,eV, respectively. In the non-relativistic case only a $\delta L_i$ parallel to the light propagation axis is induced which is odd in the helicity, and no $\delta S_i$ is induced, exemplifying that the orbital response is the primary non-relativistic one, whereas the spin response is generated through SOC, as was also shown in~\cite{Berritta_2016}. Similarly, for the antiferromagnetic case, only a component parallel to the light propagation axis is induced which is odd in the helicity and remains unchanged under different polarization flavors. While this is the case also for the induced components $\delta L_x$ and $\delta L_y$ which are transverse to the magnetization in the ferromagnetic case, the situation drastically changes for the induced component $\delta L_z$ parallel to the magnetization, as discussed earlier. We note that a tiny, odd in the helicity, $\delta L_x$ or $\delta L_y$ is additionally induced when light is rotating in a plane containing ferromagnetic magnetization. On the other hand, the induced $\delta L_x$ and $\delta L_y$, developing normal to the polarization plane, serve as a non-relativistic ``background" which is independent of the magnetization, with features due to the crystal structure driving the effect over larger regions in energy. The additional band-splittings induced by SOC result in the IFE exhibiting more features with band filling in the relativistic scenario. Remarkably, while the relativistic antiferromagnetic and non-relativistic ferromagnetic cases in principle have similar to each other behavior in energy, a larger signal arises in the antiferromagnetic case by the virtue of flatter bands (see also the discussion for Hf and Pt below). Lastly, we note that a similar behavior has been observed for the in-plane spin IFE   in PT-symmetric Mn$_2$Au, however, in the latter case additional  out-of-plane moments arise due to linearly polarized light as a result of broken by the magnetization inversion symmetry~\cite{Merte_Mn2Au}.

\begin{figure*}[t!]
\begin{center}
\rotatebox{0}{\includegraphics [width=0.85\linewidth]{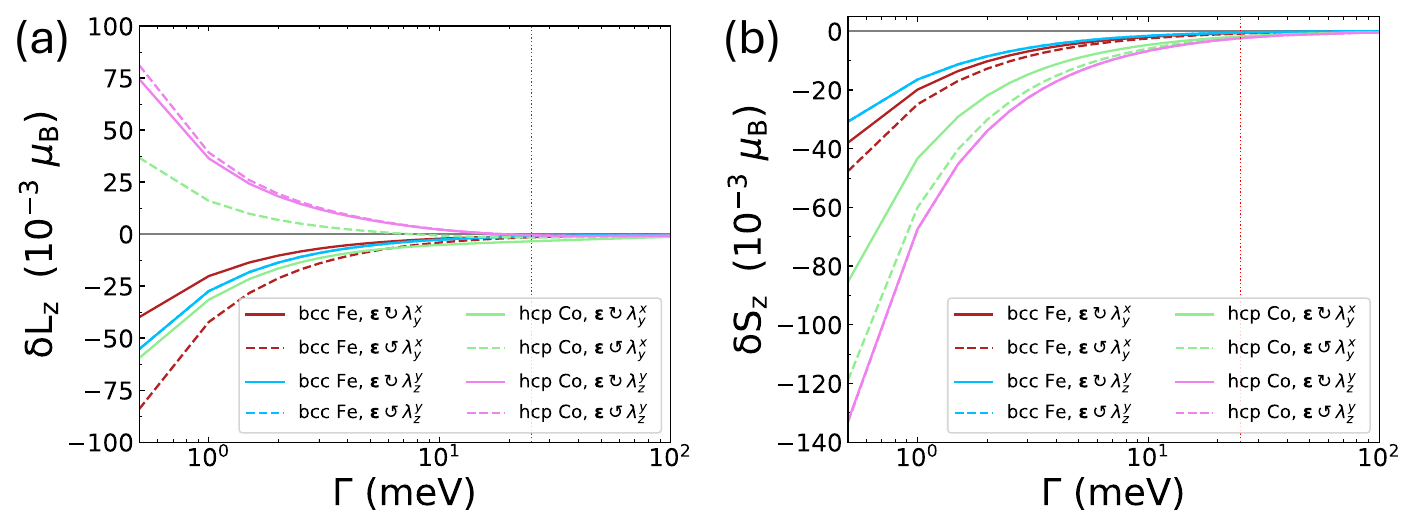}}
\end{center}
\caption{{\bf Anisotropy of light-induced magnetism in ferromagnetic transition metals}. (a-b) Light-induced orbital $\delta L_z$ (a) and spin $\delta S_z$ (b) magnetic moments in relation to the lifetime broadening for ferromagnetic bcc Fe and hcp Co. Light is circularly polarized in the $xy$-plane (brown curves for bcc Fe and light green curves for hcp Co) or in the $yz$-plane (light blue curves for bcc Fe and light purple curves for hcp Co). Both right-handed (solid lines) and left-handed (dashed lines) polarizations are displayed. The ferromagnetic magnetization is along the $z$-axis. The light frequency is set at $\hbar\omega=1.55$\,eV.}
\label{Fig3}
\end{figure*}

Next, we focus on the case of $5d$ transition metals where in Fig.~\ref{Fig2}(a-b) we explore the relation of $\delta L_z$ and $\delta S_z$, respectively, to the band filling, for light circularly polarized  in the $xy$ plane and the frequency of $\hbar\omega=1.55$\,eV. When going from group IV (hcp Hf) to XI (fcc Au) we observe a smooth variation of $\delta L_z$ from positive to negative values, as well as nicely shaped plateaus, where $\delta L_z$ remains relatively robust in a wide energy region, for hcp Hf, bcc Ta, bcc W, fcc Ir, and fcc Pt. We note that such plateaus are often characteristic of orbital effects, as witnessed for example in orbital Hall insulators~\cite{Canonico_2020, Cysne_2021, Zeer_2022} and orbital Rashba systems~\cite{Adamantopoulos_2024}. On the contrary, $\delta S_z$ exhibits a very erratic behavior with strong variations for each material, at the same time being much smaller in magnitude than $\delta L_z$. The above observation is a clear manifestation of how differently the orbital and spin degrees of freedom behave under light excitation, with the orbital angular momentum having its origin in intrinsic structural parameters as manifested in the crystal field splitting, whereas the spin angular momentum is more sensitive to finer details of the electronic structure mediated by SOC. Indeed, the light-induced spin and orbital moments exhibit similarly erratic behavior with band filling once the frequency of the light is drastically reduced to reach the range of spin-orbit interaction, see Fig.~\ref{Fig4_suppl} of the SM for $\hbar\omega=0.25$\,eV.
\vspace{0.2cm}

\noindent{\bf Anatomy of IFE in $k$-space}

\noindent Among the $5d$ transition metals, hcp Hf and fcc Pt exhibit the largest computed moments at the Fermi energy for $\hbar\omega=1.55$\,eV, with the corresponding values of $\delta L_z=9.1\cdot10^{-3}\mu_{\mathrm{B}}$, $\delta S_z=-0.8\cdot10^{-3}\mu_{\mathrm{B}}$ for Hf, and $\delta L_z=-7.4\cdot10^{-3}\mu_{\mathrm{B}}$, $\delta S_z=-1.1\cdot10^{-3}\mu_{\mathrm{B}}$ for Pt. Therefore, we select these two materials and explore the behavior of their light-induced moments in reciprocal space. We present the band-resolved $\delta L_z$ and $\delta S_z$ for hcp Hf in Fig.~\ref{Fig5}(a-b), as well as band-resolved $\delta L_z$ and $\delta S_z$ for fcc Pt in Fig.~\ref{Fig5}(c-d). For the case of Hf, transitions along the bands near the A-point are the main source of $\delta L_z$ and $\delta S_z$. Light-induced $\delta L_z$ consists of hotspot-like negative contributions and secondary, but extended over energy and $k$-space consistently positive contributions, which are hardly visible. Overall, however, the latter lead to a large positive orbital integrated response. On the other hand, $\delta S_z$ in Hf presents a much richer anatomy with both negative and positive contributions at times emerging within the same band in the narrow neighboring regions of the reciprocal space.

For the case of Pt, $\delta L_z$ and $\delta S_z$ arise from transitions close to X and L high-symmetry points, with both originating in roughly the same regions of $(\mathcal{E},k)$-space, but often having an opposite sign to each other. Note that the bands in Pt are much more dispersive in the considered energy window, which results in an effective reduction of the regions in $(\mathcal{E},k)$-space which contribute to the spin and orbital response alike. This is in contrast to Hf, where much flatter bands reside above and below the Fermi energy within the energy window of the laser pulse, providing significant integrated, albeit very small locally, contributions. Moreover, in fcc Pt the Fermi energy cuts through the band edges of the $d$-states, where the effect of spin-orbit interaction is the strongest, which explains the emergence of strong hotspot-like contributions with a clear correlation in the magnitude of spin and orbital response  and the resultant similar behavior of $\delta L_z$ and $\delta S_z$ with band filling around the Fermi energy.

\begin{figure*}[t!]
\begin{center}
\rotatebox{0}{\includegraphics [width=0.85\linewidth]{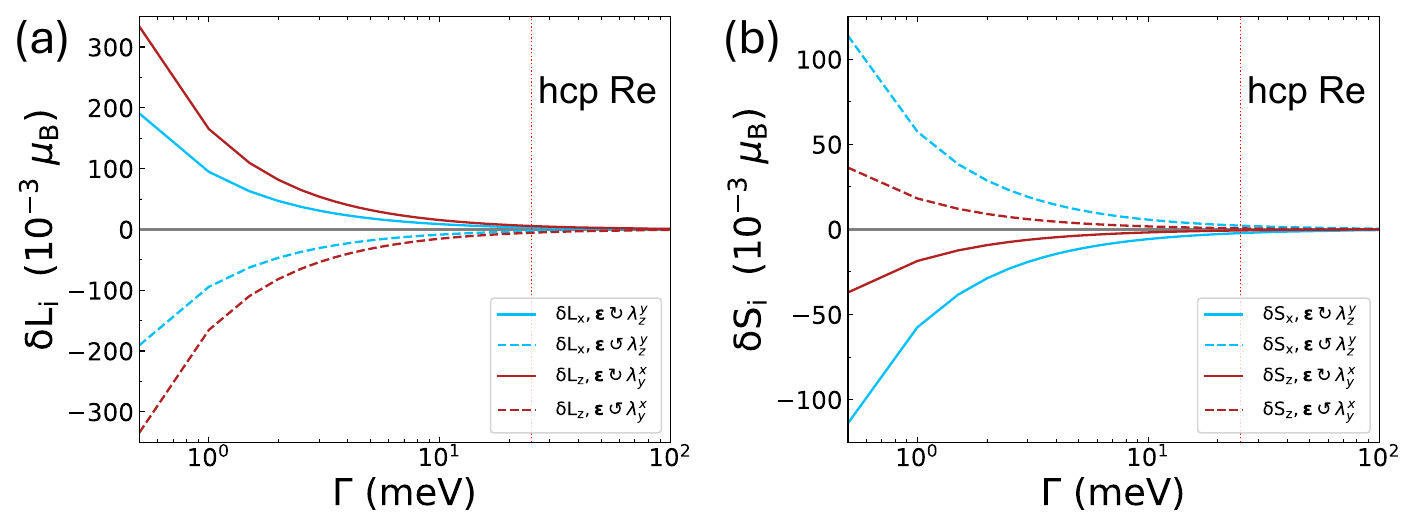}}
\end{center}
\caption{{\bf Crystalline anisotropy of light-induced magnetism in nonmagnetic transition metals}. (a-b) Light-induced orbital $\delta L$ (a) and spin $\delta S$ (c) magnetic moments in relation to the lifetime broadening for nonmagnetic hcp Re. The induced magnetic moments along the $x$-axis and along the $z$-axis are shown. Light is considered to be circularly polarized in the $xy$-plane (brown curves) or in the $yz$-plane (light blue curves). Both right-handed (solid lines) and left-handed (dashed lines) polarizations are displayed. The light frequency is set at $\hbar\omega=1.55$\,eV.}
\label{Fig4}
\end{figure*}

We further scrutinize the reciprocal space distribution of $\delta L_z$ and $\delta S_z$, shown for hcp Hf and fcc Pt in Fig.~\ref{Fig5}(e-h). For both materials $\delta L_z$ distributions consist of large uniform areas of either positive or negative sign. On the contrary, $\delta S_z$ distributions are much finer and richer in details with more areas of opposite sign, consistent with the picture we drew above from the band-resolved analysis. A similar behavior of the light-induced magnetism in reciprocal space has been recently reported for rutile altermagnets~\cite{Adamantopoulos_npj_2024}. We also observe that, as discussed above, while for Hf the contributions are well spread throughout the Brillouin zone, for Pt the spin and orbital contributions are located at the edges of the considered $k_z-k_y$ plane. Overall, this fact indicates that the microscopic behavior of light-induced magnetism varies strongly among transition metals and crucially depends on the crystal structure and position of the Fermi level with respect to the states split by the crystal field and spin-obit interaction. 
\vspace{0.2cm}

\noindent {\bf Anisotropy of light-induced magnetism}

\noindent We first address the anisotropy of light-induced magnetism with respect to  the magnetization by examining the response of magnetic elements under different flavors of circular polarization. In Fig.~\ref{Fig3}(a-b) we present the computed $\delta L_z$ and $\delta S_z$ in relation to scattering lifetime $\Gamma$, for the cases of magnetic bcc Fe and hcp Co under excitation by light which is circularly polarized in the $xy$ or $yz$ planes at $\hbar\omega=1.55$\,eV. As we have already seen in Fig.~\ref{Fig1} for light polarized in the $xy$-plane, the responses behave differently for right (solid lines) or left (dashed lines) polarization. This difference is more pronounced for $\delta L_z$, for which we even observe a change of sign for Co. Surprisingly, the situation is drastically different for the case of $yz$-polarization. In this case, for bcc Fe the response is even in the helicity, while for hcp Co $\delta L_z$ is almost even and $\delta S_z$ is perfectly even in the helicity. Such different behavior with respect to the light helicity between the two magnetic elements can be traced back to the additional anisotropy originating in the crystal structure itself $-$ an effect which we discuss below. Notably, we witness a highly non-linear behavior with respect to $\Gamma$, with the response reaching colossal values in the clean limit.
For example, $\delta L_z$ and $\delta S_z$ can reach as much as $40\cdot10^{-3}\mu_{\mathrm{B}}$ and $70\cdot10^{-3}\mu_{\mathrm{B}}$ in Co, respectively, given the scattering lifetime of 1\,meV.  Besides the potential tunability of the inverse Faraday effect by the degree of the disorder of the samples, one key message that we extract from our observations is that when comparing the values for laser-induced magnetic moments obtained with different methods, special care has to be taken since the implementations of disorder effects, even within a simple constant broadening model, may differ among various approaches.


Finally, we analyze the anisotropy that the crystal structure induces in the light-induced magnetism. We select the case of hcp Re and present in Fig.~\ref{Fig4}(a-b) the $x$ and $z$ components of $\delta L$ and $\delta S$, respectively, in relation to $\Gamma$. Due to the inequivalence of $x$ and $z$ axes in the hcp structure, the $\delta L_z$ which originates in light circularly polarized  in the $xy$-plane (light blue line) differs from the $\delta L_x$ for the case of the $yz$ polarization (brown line). The situation is similar for the case of $\delta S_z$ and $\delta S_x$. On the  contrary, as presented in Fig.~\ref{Fig3_suppl} of the SM, in the case of fcc Pt we have a perfect match between $\delta L_z$ and $\delta L_x$ (or $\delta S_z$ and $\delta S_x$) when changing the plane of circular polarization because the $x$ and $z$ axes are equivalent in the fcc structure. The situation is similar for the $y$ component when light is circularly polarized in the $xz$-plane. As also expected from the symmetry of the crystal structure, we confirm that the light-induced moments are perfectly odd in the helicity for the nonmagnetic elements, which is not the case for the magnetic elements, see  Figs.~\ref{Fig4} and~\ref{Fig3_suppl} of the SM~\cite{Berritta_2016}.
\vspace{0.2cm}


\noindent {\bf Discussion}

\noindent The main goal of our work is to showcase the importance of the orbital component of IFE and its distinct behavior from the spin counterpart. While so far it was mainly the interaction of light with the spin magnetization that was taken into consideration for the interpretation of IFE-related effects, we speculate that the orbital IFE may provide a novel way to coherently induce magnetization and manipulate the magnetic order. For example, in a recent study, different types of optical torques that may arise in ferromagnetic layers were interpreted in terms of the light-induced orbital moment and its interaction with the magnetization through the spin-orbit interaction~\cite{Nukui_2024}. Since it is known that current-induced orbital accumulation and orbital torques exhibit a long-range behavior~\cite{Choi_2023, Go_2023, Bose_2023, Hayashi_2023} due to a characteristic small orbital decay, a question arises whether similar behavior can be exhibited by  optical torques caused by the orbital IFE. The emergence of long-ranged orbital IFE would come as no surprise given the fact that several recent experiments reported that laser excitation can drive long-range ballistic orbital currents resulting in THz emission~\cite{Seifert_2023, Wang_2023, Xu_2024, Liu_2024}. 

Although in our treatment the effect of light enters only as a perturbation by the electric field of the original ground state Hamiltonian, it is common to utilize the angular momentum of light in order to understand the interaction with the magnetic order by the means of transfer of angular momentum. The spin of light through the helicity of circularly polarized pulses plays a crucial role in helicity-dependent all-optical switching scenarios. On the other hand, there is a strong recent interest in utilizing the orbital angular momentum of light via irradiation of matter with e.g. vortex beams or twisted light, in order to probe the magnetization~\cite{Sirenko_2019}, generate photocurrents~\cite{Ji_2020}, drive ultrafast demagnetization~\cite{Prinz_2022}, and induce IFE~\cite{Ali_2010, Karakhanyan_2022}. Therefore, it is imperative to treat both spin and orbital degrees of freedom on equal footing when exploring the light-matter interaction. This will not only trigger further advances in the field of ultrafast magnetism and THz spintronics, but also enable a  transition to the novel field of attosecond spintronics~\cite{Siegrist_2019, Neufeld_2023}.
\vspace{0.2cm}

\noindent {\bf Method.}
In this work we calculate the first-principles electronic structures of $3d$, $4d$ and $5d$ transition metals by using the full-potential linearized augmented plane wave \texttt{FLEUR} code~\cite{fleurCode}. We describe exchange and correlation effects by using the non-relativistic PBE~\cite{pbe} functional, while relativistic effects are described by the second-variation scheme~\cite{SOC_2nd_var}. The parameters of our first-principles calculations, i.e. lattice constants, muffin-tin radii, plane-wave cutoffs, etc. are taken from Table I of Ref.~\cite{Go_2024}. 
	
Next, we construct maximally-localized Wannier functions (MLWFs) by employing the Wannier90 code~\cite{Pizzi2020} and its interface with the \texttt{FLEUR} code~\cite{code_fleurWann}. Similarly to~\cite{Go_2024}, we choose $s$, $p$ and $d$ orbitals for the initial projections and disentagle 18 MLWFs out of 36 Bloch states within a frozen window of 5.0\,eV above the Fermi energy for each atom in the unit cell.

Lastly, at a post-processing step, we calculate the laser-induced orbital and spin magnetizations according to the Keldysh formalism~\cite{freimuth_2016, Adamantopoulos_npj_2024}:
\begin{align}\label{eq:keldysh}	\delta\mathcal{O}_{i}=-\frac{\hbar a_{0}^{3} I}{2 c}\frac{\mathcal{E}_{\mathrm{H}}}{\left( \hbar \omega\right)^{2}} \operatorname{Im} \sum_{j k} \epsilon_{j} \epsilon_{k}^{*} \varphi_{i j k},
\end{align}
where $\mathcal{O}_{i}$ is the $i$-th component of either the orbital angular momentum operator $L_{i}$ or of the spin operator $S_{i}$. Moreover, $a_0 = 4\pi \epsilon_0 \hbar^2 /(m_e e^2)$ is the Bohr's radius, $I = \epsilon_0 c E_0^2/2$ is the intensity of the pulse, $\epsilon_0$ is the vacuum permittivity, $m_e$ is the electron mass, $e$ is the elementary charge,  $\hbar$ is the reduced Planck constant, $c$ is the light velocity, $\mathcal{E}_H=e^2/(4\pi \epsilon_0 a_0)$ is the Hartree energy, and $\epsilon_j$ is the $j$-th component of the polarization vector of the pulse. For example, we describe right/left-handedly polarized light in the $xy$-plane as $\epsilon=(1,\pm i,0)/\sqrt{2}$, respectively, and define its propagation vector to lie along the normal to the polarization plane. A detailed form of the tensor $\varphi_{ijk}$ can be seen in Eq.(14) of Ref.~\cite{freimuth_2016}. For the orbital response, the prefactor in Eq.~\eqref{eq:keldysh} must be multiplied by an additional factor of 2.
A 128$\times$128$\times$128 interpolation $k$-mesh is sufficient to obtain well-converged results. In all calculations the lifetime broadening $\Gamma$ was set at 25\,meV, the light frequency $\hbar\omega$ at 0.25\,eV and 1.55\,eV, the intensity of light at 10\,GW/cm$^2$, and we covered an energy region of $[-2.5, 2.5]$\,eV around the Fermi energy level $\mathcal{E}_F$.


\noindent {\bf Acknowledgements.} We thank Frank Freimuth and Maximilian Merte for discussions. This work was supported by the Deutsche Forschungsgemeinschaft (DFG, German Research Foundation) $-$ TRR 173/3 $-$ 268565370 (project A11) and by the K. and A. Wallenberg Foundation (Grants No. 2022.0079 and 2023.0336).
We acknowledge support from the EIC Pathfinder OPEN grant 101129641 “OBELIX”.
We  also gratefully acknowledge the J\"ulich Supercomputing Centre and RWTH Aachen University for providing computational resources under projects  jiff40 and jara0062. 

\noindent {\bf Author Contributions.}
T. A. performed numerical calculations and analysed the results. T. A. and Y. M. wrote the manuscript. All authors participated in discussions of the results and reviewing of the manuscript. Y. M. conceived the idea and supervised the project.

\noindent {\bf Competing interests.} The authors declare no competing interests.

\noindent {\bf Data availability.}
The data presented in this work can be available from the corresponding author upon reasonable request.


\hbadness=99999 
\bibliography{literature}

 \cleardoublepage
 \appendix

 \onecolumngrid

 \begin{center}
     {\bf \large Supplemental Material}
 \end{center}

\begin{figure*}[ht!]
\begin{center}
\rotatebox{0}{\includegraphics [width=0.85\linewidth]{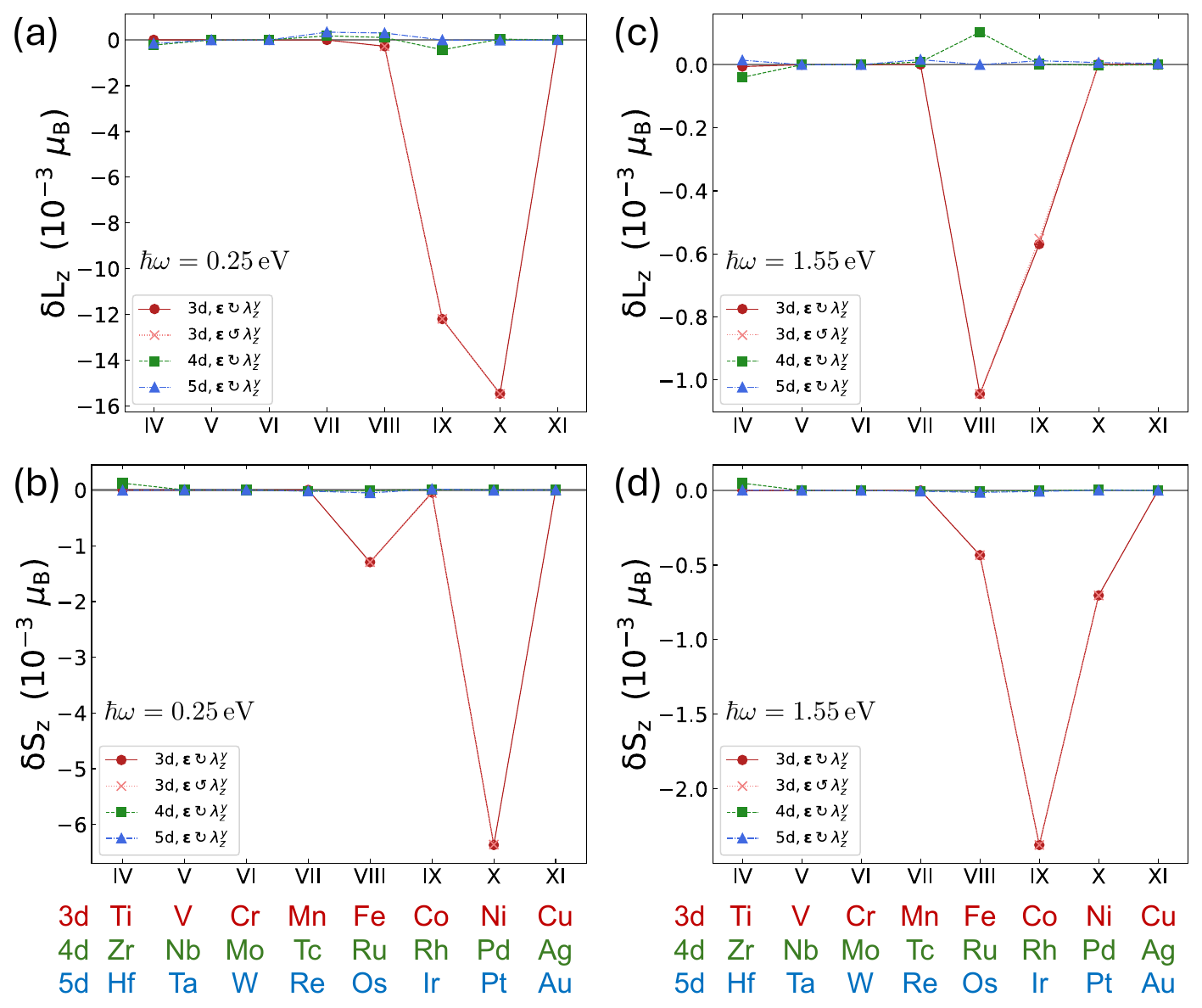}}
\end{center}
\caption{
(a-d) Light-induced orbital $\delta L_z$ (a, c) and spin $\delta S_z$ (b, d) magnetic moments in $3d$ (red circles), $4d$ (green squares) and $5d$ (blue triangles) transition metals of groups IV$-$XI. In all calculations light is considered to be circularly polarized in the $yz$-plane. For the $3d$ magnetic elements (Fe, Co, Ni) the moments that arise for left-handedly polarized light are shown with light red crosses. The light frequency is $\hbar\omega=0.25$\,eV in (a-b) and $\hbar\omega=1.55$\,eV in (c-d).}
\label{Fig1_suppl}
\end{figure*}

\begin{figure*}[ht!]
\begin{center}
\rotatebox{0}{\includegraphics [width=0.85\linewidth]{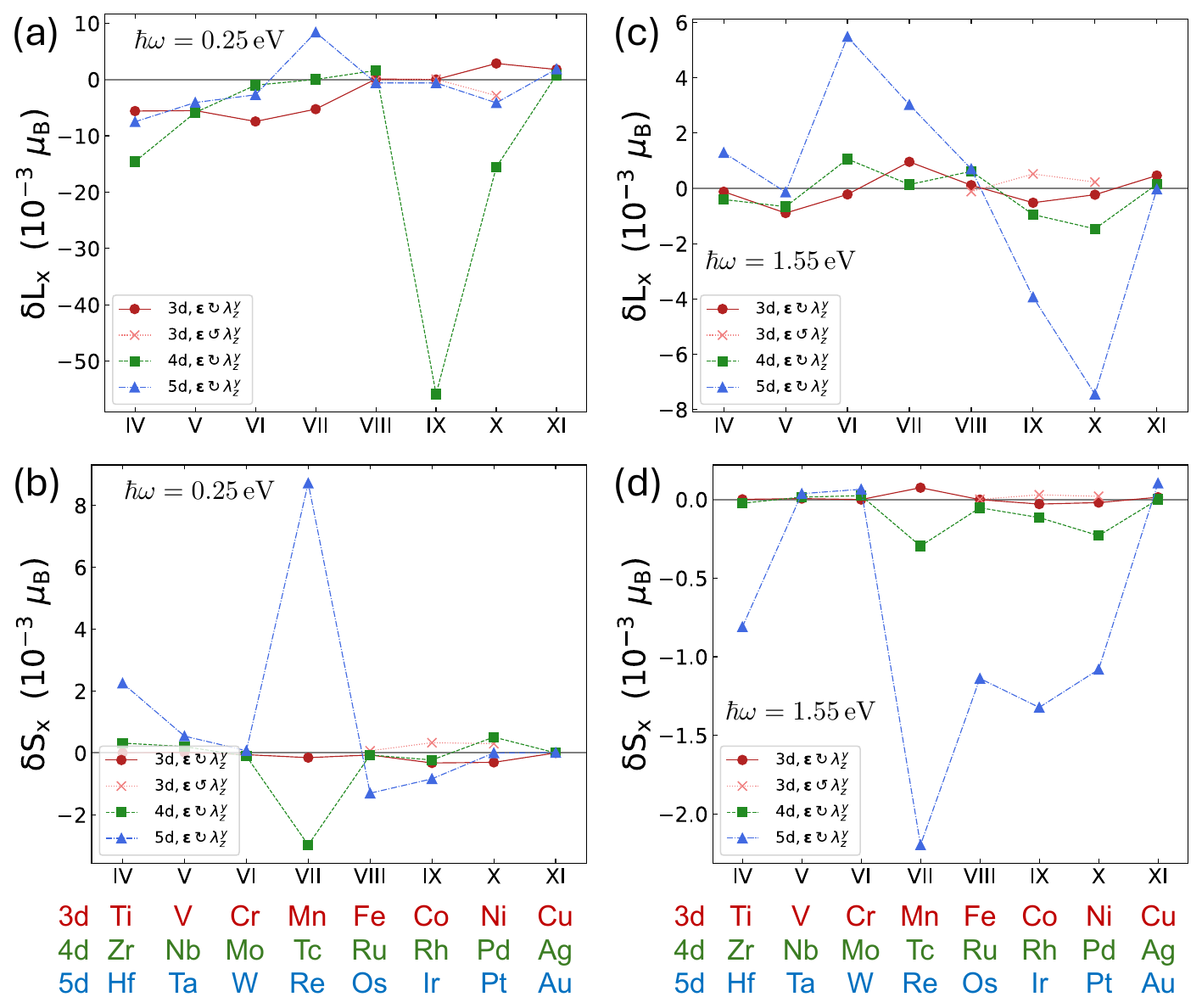}}
\end{center}
\caption{
(a-d) Light-induced orbital $\delta L_x$ (a, c) and spin $\delta S_x$ (b, d) magnetic moments in $3d$ (red circles), $4d$ (green squares) and $5d$ (blue triangles) transition metals of groups IV$-$XI. In all calculations light is considered to be circularly polarized in the $yz$-plane. For the $3d$ magnetic elements (Fe, Co, Ni) the moments that arise for left-handedly polarized light are shown with light red crosses. The light frequency is $\hbar\omega=0.25$\,eV in (a-b) and $\hbar\omega=1.55$\,eV in (c-d).}
\label{Fig2_suppl}
\end{figure*}

\begin{figure*}[t!]
\begin{center}
\rotatebox{0}{\includegraphics [width=0.95\linewidth]{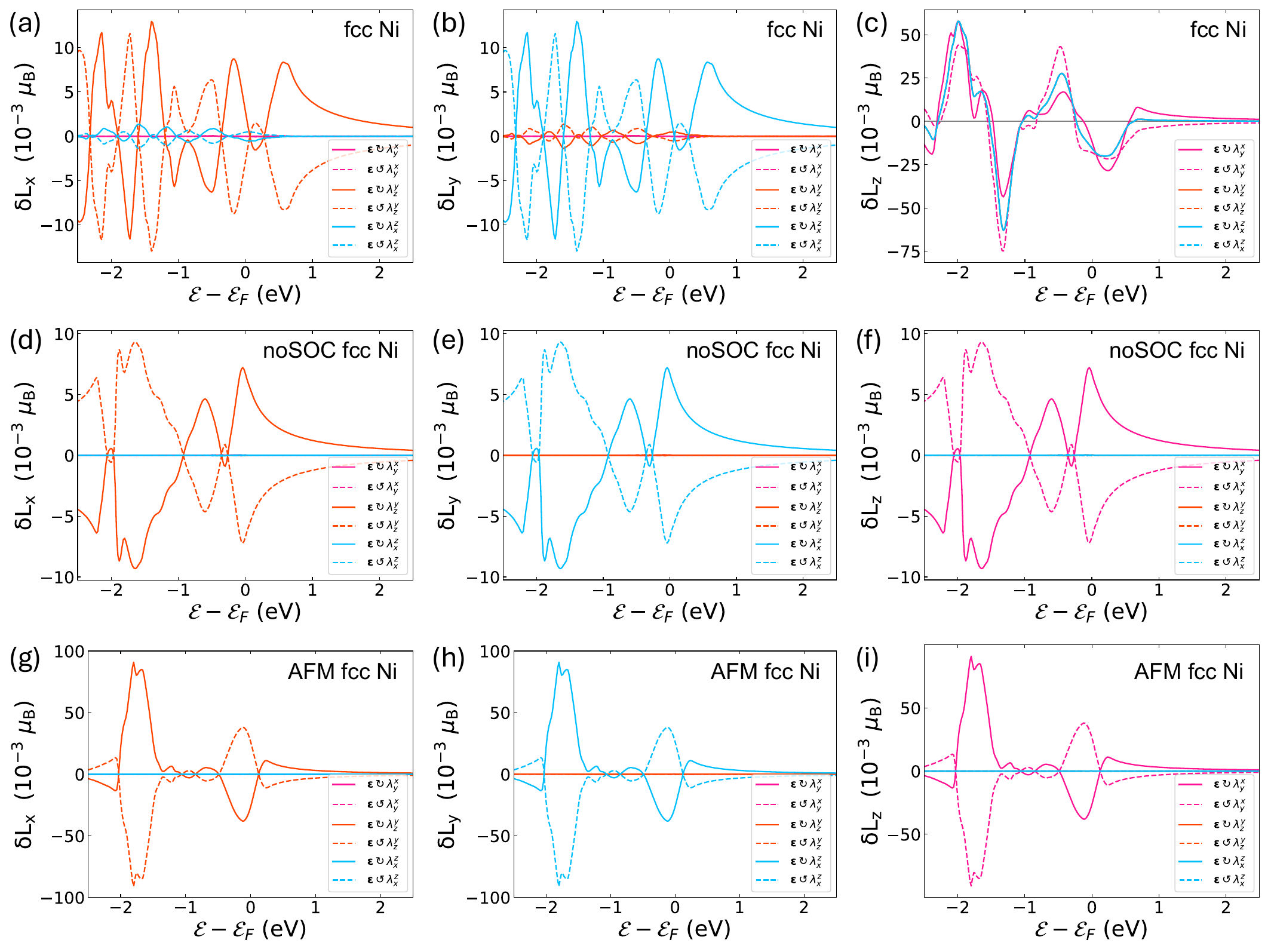}}
\end{center}
\caption{(a-i) Cartesian components of the light-induced orbital $\delta L$ magnetic moments in relation to the band filling for the cases of ferromagnetic (a-c), non-relativistic (d-f), and antiferromagnetic (g-i) fcc Ni. Light is circularly polarized in the $xy$ (pink line), $yz$ (orange line), and $zx$ (light blue line) planes, with both right-handed (solid lines) and left-handed (dashed lines) polarizations being displayed. The light frequency is set at $\hbar\omega=0.25$\,eV.}
\label{Fig5_suppl}
\end{figure*}

\begin{figure*}[t!]
\begin{center}
\rotatebox{0}{\includegraphics [width=0.95\linewidth]{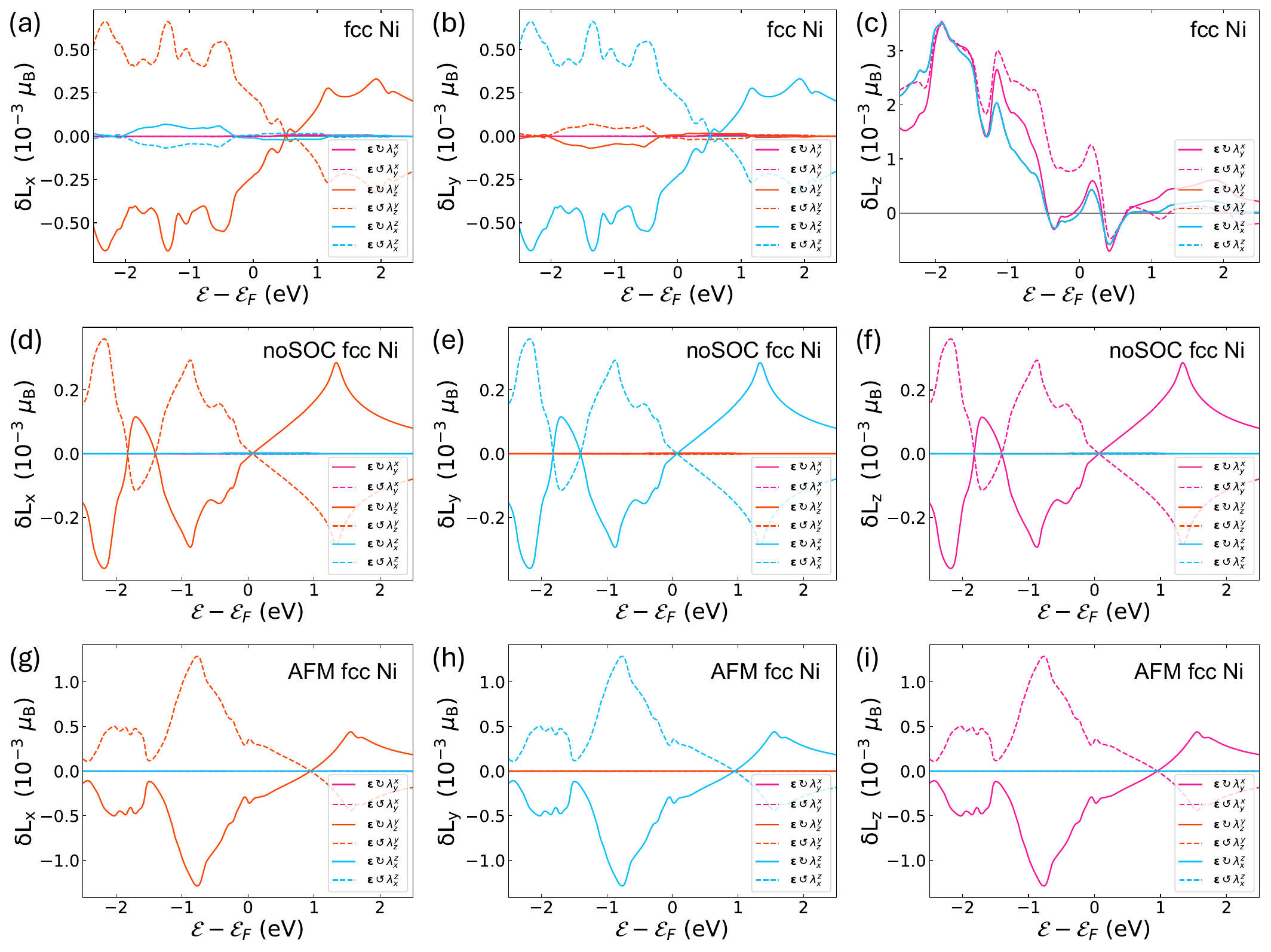}}
\end{center}
\caption{(a-i) Cartesian components of the light-induced orbital $\delta L$ magnetic moments in relation to the band filling for the cases of ferromagnetic (a-c), non-relativistic (d-f), and antiferromagnetic (g-i) fcc Ni. Light is circularly polarized in the $xy$ (pink line), $yz$ (orange line), and $zx$ (light blue line) planes, with both right-handed (solid lines) and left-handed (dashed lines) polarizations being displayed. The light frequency is set at $\hbar\omega=1.55$\,eV.}
\label{Fig6_suppl}
\end{figure*}

\begin{figure*}[t!]
\begin{center}
\rotatebox{0}{\includegraphics [width=0.85\linewidth]{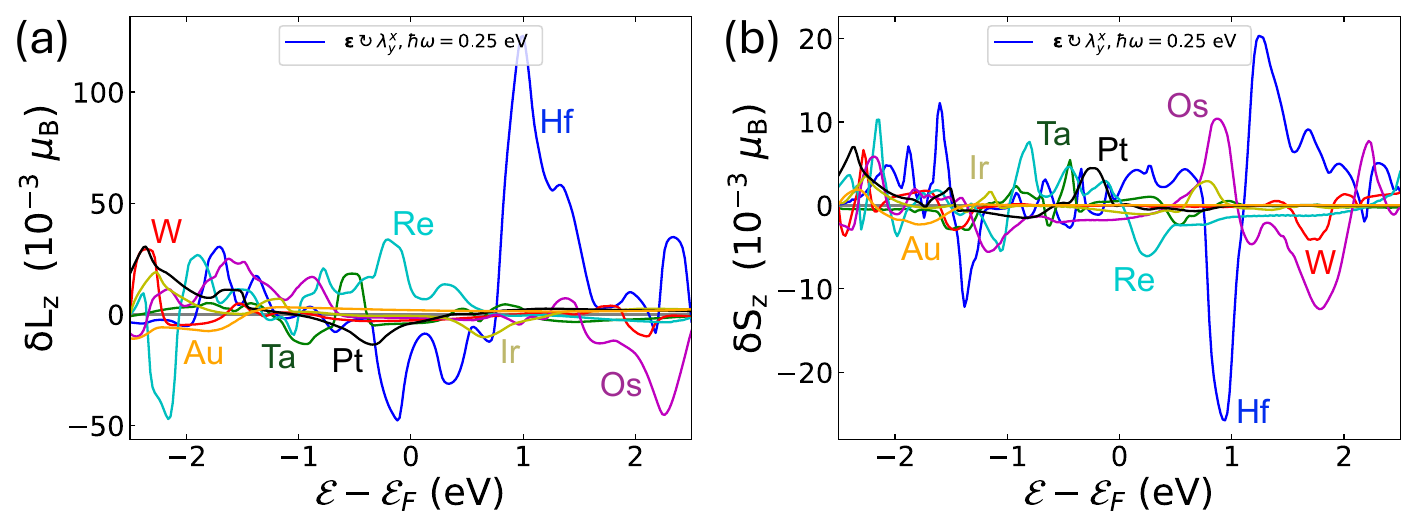}}
\end{center}
\caption{(a-b) Light-induced orbital $\delta L_z$ (a) and spin $\delta S_z$ (b) magnetic moments in relation to the band filling for the $5d$ transition metals of groups IV$-$XI. Light is circularly polarized in the $xy$-plane and the light frequency is set at $\hbar\omega=0.25$\,eV.}
\label{Fig4_suppl}
\end{figure*}

\begin{figure*}[ht!]
\begin{center}
\rotatebox{0}{\includegraphics [width=0.85\linewidth]{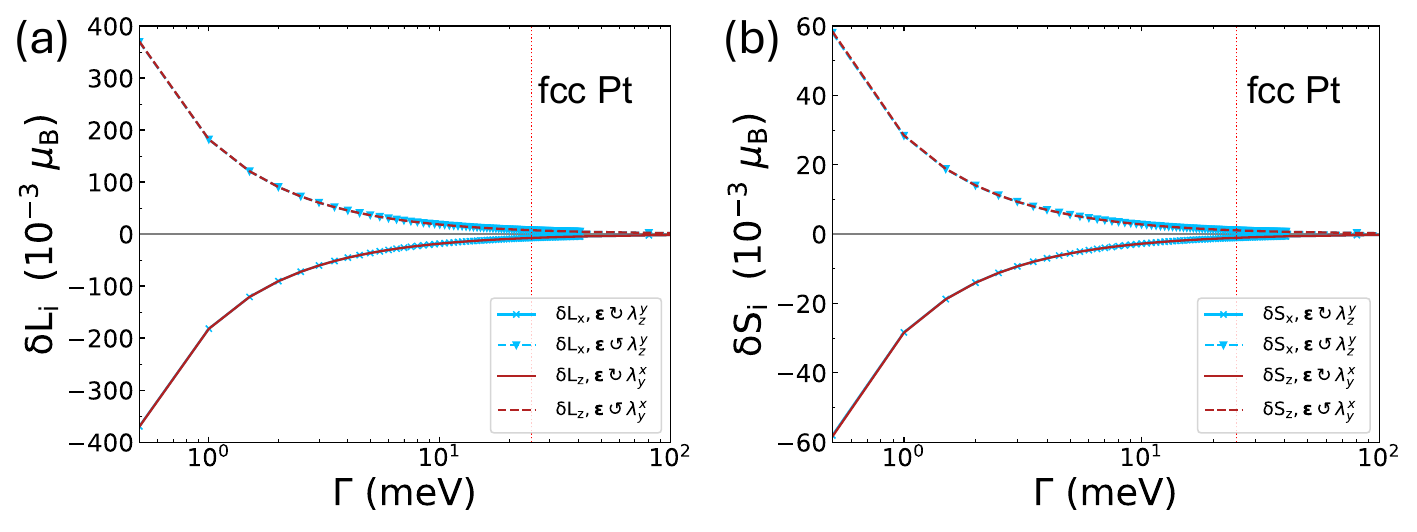}}
\end{center}
\caption{
(a-b) Light-induced orbital $\delta L$ (a) and spin $\delta S$ (c) magnetic moments in relation to the lifetime broadening for nonmagnetic fcc Pt. The induced magnetic moments along the $x$-axis and along the $z$-axis are presented. Light is considered to be circularly polarized in the $xy$-plane (brown curves) or in the $yz$-plane (light blue curves). Both right-handed (solid lines) and left-handed (dashed lines) polarizations are displayed. The light frequency is set at $\hbar\omega=1.55$\,eV.}
\label{Fig3_suppl}
\end{figure*}

\begin{table}[t!]
    \centering
    \begin{tabular}{c c c}
    \hline
    \multicolumn{3}{c}{$\hbar\omega=0.25$\,eV} \\
    \hline
   & & \\
     Material & $\delta L_z (10^{-3}\,\mu_{\mathrm{B}}/\text{unit cell})$ & $\delta S_z (10^{-3}\,\mu_{\mathrm{B}}/\text{unit cell})$ \\ [0.5ex]
     \hline \\
     hcp Ti & $-$0.6 &  0.001 \\ [0.5ex]
     bcc V  & $-$5.5 &  0.02  \\ [0.5ex]
     bcc Cr & $-$7.4 & $-$0.1   \\ [0.5ex]
     fcc Mn & $-$5.3 & $-$0.2   \\ [0.5ex]
     bcc Fe &  1.2 / 1.4 & $-$1.0 / $-$0.5   \\ [0.5ex]
     hcp Co &  2.4 / $-$9.8 &  1.8 / 3.3   \\ [0.5ex]
     fcc Ni & $-$11.6 / $-$18.0 &  $-$5.9 / $-$5.6   \\ [0.5ex]
     fcc Cu &  1.8 &  0.001 \\ [0.5ex]
     \hline \\
     hcp Zr & $-$6.3 & $-$0.1   \\ [0.5ex]
     bcc Nb & $-$5.9 &  0.2   \\ [0.5ex]
     bcc Mo & $-$1.0 & $-$0.1   \\ [0.5ex]
     hcp Tc & $-$5.5 & $-$6.0   \\ [0.5ex]
     hcp Ru & $-$7.0 &  1.0   \\ [0.5ex]
     fcc Rh &$-$55.8 & $-$0.2   \\ [0.5ex]
     fcc Pd &$-$15.6 &  0.5   \\ [0.5ex]
     fcc Ag &  0.8 &  0.001 \\ [0.5ex]
     \hline \\
     hcp Hf &$-$18.0 &  1.7   \\ [0.5ex]
     bcc Ta & $-$4.1 &  0.5   \\ [0.5ex]
     bcc W  & $-$2.7 &  0.1   \\ [0.5ex]
     hcp Re & 11.5 &  0.7   \\ [0.5ex]
     hcp Os & $-$1.5 & $-$1.7   \\ [0.5ex]
     fcc Ir & $-$0.6 & $-$0.8   \\ [0.5ex]
     fcc Pt & $-$4.1 & $-$0.002 \\ [0.5ex]
     fcc Au &  1.9 &  0.01  \\ [0.5ex]
    \hline\hline
    \end{tabular}
    \caption{Light-induced orbital $\delta L_z$ and spin $\delta S_z$ magnetic moments for the transition metals of groups IV$-$XI at the Fermi level, in units of $10^{-3}\,\mu_{\mathrm{B}}$ per unit cell. Light is circularly polarized in the $xy$-plane. For the $3d$ magnetic elements (Fe, Co, Ni) the moments that arise for both right/left-handedly polarized light are listed. The light frequency is $\hbar\omega=0.25$\,eV.}
    \label{table_1}
\end{table}

\begin{table}[t!]
    \centering
    \begin{tabular}{c c c}
    \hline
    \multicolumn{3}{c}{$\hbar\omega=1.55$\,eV} \\
    \hline
   & & \\
     Material & $\delta L_z (10^{-3}\,\mu_{\mathrm{B}}/\text{unit cell})$ & $\delta S_z (10^{-3}\,\mu_{\mathrm{B}}/\text{unit cell})$ \\ [0.5ex]
     \hline \\
     hcp Ti & $-$0.3  &  0.001 \\ [0.5ex]
     bcc V  & $-$0.9  &  0.01  \\ [0.5ex]
     bcc Cr & $-$0.2  & $-$0.0004\\ [0.5ex]
     fcc Mn &  1.0  &  0.1   \\ [0.5ex]
     bcc Fe & $-$0.9 / $-$1.5 & $-$0.6 / $-$0.8   \\ [0.5ex]
     hcp Co & $-$3.5 / $-$1.6 & $-$1.8 / $-$2.0   \\ [0.5ex]
     fcc Ni &  0.1 /  0.8 & $-$0.8 / $-$0.6   \\ [0.5ex]
     fcc Cu &  0.5  &  0.01 \\ [0.5ex]
     \hline \\
     hcp Zr & $-$0.9  & $-$0.1   \\ [0.5ex]
     bcc Nb & $-$0.7  &  0.02  \\ [0.5ex]
     bcc Mo &  1.1  &  0.02  \\ [0.5ex]
     hcp Tc &  2.2  & $-$0.2   \\ [0.5ex]
     hcp Ru &  0.3  &  0.2   \\ [0.5ex]
     fcc Rh & $-$0.9  & $-$0.1   \\ [0.5ex]
     fcc Pd & $-$1.5  & $-$0.2   \\ [0.5ex]
     fcc Ag &  0.1  &  0.0005\\ [0.5ex]
     \hline \\
     hcp Hf &  9.1  & $-$0.8   \\ [0.5ex]
     bcc Ta & $-$0.1  &  0.04  \\ [0.5ex]
     bcc W  &  5.5  &  0.1   \\ [0.5ex]
     hcp Re &  5.5  & $-$0.6   \\ [0.5ex]
     hcp Os &  1.2  &  0.5   \\ [0.5ex]
     fcc Ir & $-$3.9  & $-$1.3   \\ [0.5ex]
     fcc Pt & $-$7.4  & $-$1.1   \\ [0.5ex]
     fcc Au & $-$0.03 &  0.1   \\ [0.5ex]
    \hline\hline
    \end{tabular}
    \caption{Light-induced orbital $\delta L_z$ and spin $\delta S_z$ magnetic moments for the transition metals of groups IV$-$XI at the Fermi level, in units of $10^{-3}\,\mu_{\mathrm{B}}$ per unit cell. Light is circularly polarized in the $xy$-plane. For the $3d$ magnetic elements (Fe, Co, Ni) the moments that arise for both right/left-handedly polarized light are listed. The light frequency is $\hbar\omega=1.55$\,eV.}
    \label{table_2}
\end{table}

\end{document}

%% file: commands.tex
\newcommand{\pgi}{Peter Gr\"unberg Institut,
Forschungszentrum J\"ulich, 52425 J\"ulich, Germany}

\newcommand{\aachen}{Department of Physics, RWTH Aachen University, 52056 Aachen, Germany}

\newcommand{\mainz}{Institute of Physics, Johannes Gutenberg University Mainz, 55099 Mainz, Germany}

\newcommand{\uppsalaa}{Department of Physics and Astronomy, Uppsala University, Box 516, SE-75120 Uppsala, Sweden}

\newcommand{\uppsalab}{Wallenberg Initiative Materials Science for Sustainability, Uppsala University, SE-75120 Uppsala, Sweden}

\newcommand{\abs}{\mathrm{\abs}}